\documentclass[12pt]{JHEP3}  
\newcommand{\da}{{\dot\a}}
\newcommand{\ad}{{\dot\a}}
\newcommand{\db}{{\dot\b}}

\def\cao{\c c\~ao}

\def\ftoday{{\sl {Le \number\day \space\ifcase\month 
\or janvier\or f\'evrier\or mars\or avril\or mai
\or juin\or juillet\or ao\^ut\or septembre\or octobre
\or novembre \or d\'ecembre\fi\space \number\year}}}
\def\ptoday{{\sl {\number\day \space de\space \ifcase\month 
\or janeiro\or fevereiro\or mar{\c c}o\or abril\or maio
\or junho\or julho\or agosto\or setembro\or outubro
\or novembro \or dezembro\fi\space de\space \number\year}}}
\def\gtoday{{\sl {Den \number\day. \ifcase\month 
\or Januar\or Februar\or M\"arz\or April\or Mai
\or Juni\or Juli\or August\or September\or Oktober
\or November \or Dezember\fi\space \number\year}}}
\def\today{{\sl {\ifcase\month
\or January\or February\or March\or April\or May
\or June\or July\or August\or September\or October
\or November \or December\fi \space\number\day,\space 
                                            \number\year}}}
\renewcommand{\a}{\alpha}
\renewcommand{\b}{\beta}
\newcommand{\g}{\gamma}

\renewcommand{\d}{\delta}         

\newcommand{\e}{\varepsilon}
\newcommand{\la}{\lambda}
\newcommand{\LA}{\Lambda}
\newcommand{\m}{\mu}

\newcommand{\OM}{\Omega}
\newcommand{\p}{\psi}

\newcommand{\s}{\sigma}

\newcommand{\f}{{\phi}}
\newcommand{\F}{{\Phi}}
\newcommand{\vf}{{\varphi}}

\newcommand{\DD}{{\cal D}}

\newcommand{\FF}{{\cal F}}

\newcommand{\LL}{{\cal L}}

\newcommand{\PP}{{\cal P}}

\newcommand{\es}{\\[3mm]}

\newcommand{\sla}{\raise.15ex\hbox{$/$}\kern -.57em}
\newcommand{\Sla}{\raise.15ex\hbox{$/$}\kern -.70em}

\newcommand{\lp}{\left(}
\newcommand{\rp}{\right)}
\newcommand{\lc}{\left[}
\newcommand{\rc}{\right]}

\newcommand{\rac}{\right\}}
\newcommand{\identity}{{\bf 1\hspace{-0.4em}1}}
\newcommand{\complex}{{\kern .1em {\raise .47ex
\hbox {$\scriptscriptstyle |$}}
    \kern -.4em {\rm C}}}
\newcommand{\real}{{{\rm I} \kern -.19em {\rm R}}}
\newcommand{\rational}{{\kern .1em {\raise .47ex
\hbox{$\scripscriptstyle |$}}
    \kern -.35em {\rm Q}}}
\renewcommand{\natural}{{\vrule height 1.6ex width
.05em depth 0ex \kern -.35em {\rm N}}}

\newcommand{\tr}{{\rm {Tr} \,}}

\newcommand{\pa}{\partial}

\newcommand{\dfrac}[2]{{\displaystyle{\frac{#1}{#2}}}}
\newcommand{\dsum}[2]{\displaystyle{\sum_{#1}^{#2}}}

\newcommand{\twiddle}{\lower.9ex\rlap{$\kern -.1em\scriptstyle\sim$}}

\newcommand{\equ}[1]{(\ref{#1})}
\newcommand{\eq}{\begin{equation}}
\newcommand{\eqn}[1]{\label{#1}\end{equation}}
\newcommand{\ba}{\begin{array}}
\newcommand{\ea}{\end{array}}

\title{Induced mass in $N=2$ super Yang-Mills theories}
\author{Sortelano Araujo Diniz$^{\rm{(a,b)}\;*}$ and 
Olivier Piguet$^{\rm{(c)}}$
\thanks{Supported by the Conselho Nacional de
Desenvolvimento Cient\'{\i}fico e Tecnol\'{o}gico CNPq -- Brazil.}
\\
$^{\rm{(a)}}$ Centro Brasileiro de Pesquisas F\'\i sicas (CBPF),\\
Coordena\c c\~ao de Teoria de Campos e Part\'\i culas (CCP),\\
Rua Dr. Xavier Sigaud 150 - 22290-180 - Rio de Janeiro - RJ - Brazil\\
$^{\rm{(b)}}$ Universidade Vale do Rio Doce\\
Faculdade de Engenharia -- FAENG Campus II\\
 35020-220 -  Capim, Governador Valadares  - MG - Brazil.\\
$^{\rm{(c)}}$ Universidade Federal do Esp\'\i rito Santo (UFES),\\
Departamento de F\'\i sica, \\
Campus Universit\'ario de Goiabeiras
- 29060-900 - Vit\'oria - ES - Brazil.\\
E-mails: \email{sortelano@univale.br}, \email{piguet@cce.ufes.br} }

\preprint{hep-th/0211039\\
UFES-DF-OP2002/1}
\keywords{Extended supersymmetry, Gauge theories;
PACS: 11.30.PB, 11.15.-q}

\abstract{The masses of the matter fields of $N=2$ Super-Yang-Mills 
theories can be defined as parameters of deformed supersymmetry 
transformations. The formulation used involves central charges for
the matter fields. The explicit form of the deformed supersymmetry
transformations and of the invariant Lagrangian in presence of the gauge
supermultiplet are constructed.
This works generalizes a former one, due to the same authors, which
presented the free matter case.}

\begin{document}

\section{Introduction}

As is well known~\cite{Wess,gir-gri}, a superspace description 
of supersymmetric 
theories has to be formulated in terms of {\it constrained} 
superfields. In the case of gauge theories with
$N=2$ extended supersymmetry, the
constraints are those on torsion and curvature of the gauge 
superfield~\cite{gsw,hin,ghh,aghh} and on the matter 
superfields (Fayet hypermultiplets)~\cite{fay,West,soh,gai}, 
inclusive on their central charge dependence. 
Our aim is to show that a generalization
of the Sohnius central charge constraint~\cite{soh} for the
hypermultiplet generates a mass for the corresponding matter
particles, this mass appearing as a parameter $\LA$ in the generalized 
central charge 
constraint and in the resulting deformed supersymmetry transformations
of the component fields. 
A preliminary version of this work~\cite{I} dealt
with the case of the free hypermultiplet only. 

We shall extend here the
construction to the general case of matter hypermultiplets coupled to
gauge fields. We shall present the (deformed) supersymmetry
transformations of the component fields, as well as the construction 
of the invariant action, using for the matter part an algorithm due to 
Hasler~\cite{Hasler}. 
The gauge supermultiplet component field content corresponds to
a gauge of the Wess-Zumino type, the gauge degree of freedoms being reduced
to those of a usual Yang-Mills theory.

Our result reproduces in a purely algebraic way that 
of~\cite{constant-f-str,constant-f-str'}. The latter indeed showed that
the masses of the matter particles
may be generated by coupling the 
matter fields to a constant Abelian
super-Yang-Mills field strength, the values of the masses
being proportional to the value of this field 
strength. 
Thus, the mass generation via a generalized central charge constraint 
as proposed in the present paper offers an alternative way of
generating the masses, with the parameter $\LA$ replacing the constant
field strength. The interest of this alternative way is that it 
appears more natural, being purely algebraic.

Independent of these considerations, the explicit construction of 
the theory with central charge dependent matter supermultiplets
as performed here, presents its own interest and, in the best of 
our knowledge, has not
yet been shown in the literature.

The plan of the paper is the following. After recalling some
basic notions for $n=2$ superspace in Section 2
and reviewing the implementation of the
generalized Sohnius constraint in the free 
case~\cite{I} in Section 3, 
we consider the case of the coupling with a
gauge supermultiplet in Section 4. Our conclusions are presented in
Section 5.

\section{$N=2$ Central charge superspace}

Flat $N=2$ superspace with central 
charge\footnote{Our notations and conventions are given in the
Appendix.}~\cite{soh,gai,hin,ghh,aghh}
will be described by the 
coordinates $\{{X}^{A}\}$ = $\{x^{a},\theta_{\alpha}^{i},
\bar{\theta}_{i\dot{\alpha}},z,\bar{z}\}$ , respectively
the space-time coordinates $x^a$, a complex
Weyl spinor - isospinor $\theta_\a^i$ and a complex central charge $z$.
The spinor coordinates $\theta$ are Grassmann
(i.e. anticommuting or ''fermionic'') numbers, the remaining ones are ordinary
(i.e. commuting or ''bosonic'') numbers, so the manifold coordinates satisfy
the (anti)commutation rules:
\begin{equation}
{X}^{A}{X}^{B}{=(-)}^{ab}{X}^{B}{X}^{A}%
\label{grading}
\end{equation}
where the grading $a=0$ if ${X}^{A}$ is bosonic, and $a=1$ in the
fermionic case.


$N=2$ supersymmetry is
defined by the Wess-Zumino superalgebra~\cite{West,Wess}
\begin{equation}
(\PP_{A},\PP_{B}\}=T_{AB}^{C}\PP_{C}\ ,\label{s-alg}%
\end{equation}
where $\PP_{A}$ = $\{P_{a},\,Q_{\alpha}^{i},\,\bar{Q}_{i\dot{\alpha}%
},\,Z,\,\bar{Z}\}$ is the set of infinitesimal generators: the translations
$P_{a}$ ($a=0,\cdots,3$), the supersymmetry generator $Q_{\alpha}^{i}$,
its hermitian conjugate
 $\bar{Q}{}_{i{\dot{\alpha}}}$
($\alpha$ and ${\dot{\alpha}}= 1,2=$ spin indices;
$\ i=1,2=SU(2)$ (isospin) index) and 
the complex central charge generator 
$Z$. Under Lorentz transformations, $P_{a}$ transforms as a
vector; $Q$ and $\bar{Q}$  as Weyl spinors,
respectively in the $(\frac{1}{2},0)$ and $(0,\frac{1}{2})$ representations;
$Z$ and $\bar{Z}$ transform as scalars. Moreover $Q$ and $\bar{Q}$ transform
as doublets of the isospin group SU(2), the remaining generators being 
singlets.

The generators $P$, $Z$ and $\bar{Z}$ are bosonic, whereas $Q$ and $\bar{Q}$
are fermionic. Accordingly, the bracket $(\cdot,\cdot\}$ in the l.h.s.
of (\ref{s-alg}) is an anticommutator if both entries are fermionic, and a
commutator otherwise.

Finally, the structure constants of the superalgebra (\ref{s-alg}) -- the
``torsions'' -- are given by:
\begin{equation}
T_\a^i{\,}_\b^j{\,}^z=2i\varepsilon^{ij}\varepsilon_{\alpha\beta}\ ,\quad
T_\da^i{\,}_\db^j{\,}^{\bar z}=2i\varepsilon^{ij}
\varepsilon_{\dot{\alpha}\dot{\beta}}\ ,\quad 
T_\a^i{\,}_\db^j{\,}^a=-2i\varepsilon
^{ij}\sigma_{\alpha\dot{\beta}}^{a}\ ,
\label{torsions}\end{equation}
all the other torsion coefficients vanishing.

Representations of the  Wess-Zumino superalgebra (\ref{s-alg}) 
are defined as superfields. A superfield ${\phi}({X})$
is a function in superspace transforming
under the generators of the superalgebra as follows:
\begin{equation}%
\begin{array}{ll}
P_{a}{\phi}={\partial}_{a}{\phi}\ , & \\[3mm]
Q_{\alpha}^{i}\phi=\left(  \partial_{\alpha}^{i}-i\sigma_{\alpha\dot{\beta}
}^{a}{\bar{\theta}}^{i{\dot{\beta}}} \pa_a
+\theta_{\alpha}^{i}\partial_{z}\right)  {\phi}\ ,\quad &
\bar{Q}{}_{i{\dot{\alpha}}}{\phi}=\left(  -{{\bar{\partial}}}_{i\dot{\alpha}%
}+i\theta_{i}^{\alpha}\sigma_{\alpha\dot{\alpha}}^{a} \pa_a
-{\bar{\theta}}_{i{\dot{\alpha}}}\pa_{\bar z}\right)  {\phi}\ ,\\[3mm]
Z{\phi}={\partial}_{z}{\phi}\ ,  &
\bar{Z}{\phi}={\partial}_{\bar{z}}{\phi}\ .
\end{array}
\label{transf-s-field}%
\end{equation}
with $\pa_a=\partial/\partial x^{a}$\,,
$\pa_\a^i=\partial/\partial\theta_i^\a$\,,
$\bar{\partial}_{i\dot{\alpha}}=
\partial/\partial{\bar{\theta}}^{i\da}$\,, etc.
This provides the superfield representation of the superalgebra (\ref{s-alg}).

The covariant derivatives $D_{A}$ are superspace derivatives defined such that
$D_{A}{\phi}$ transforms in the same way as the superfield ${\phi}$ itself.
They are given by
\begin{equation}%
\begin{array}{ll}
D_{a}{\phi}={\partial}_{a}{\phi}\ , & \\[3mm]
D_{\alpha}^{i}\phi=\left(  \partial_{\alpha}^{i}+i\sigma_{\alpha\dot{\beta}
}^{a}{\bar{\theta}}^{i{\dot{\beta}}} \pa_a
-\theta_{\alpha}^{i}\partial_{z}\right)  {\phi}\ ,\quad &
\bar{D}{}_{i{\dot{\alpha}}}{\phi}=\left(  -{{\bar{\partial}}}_{i\dot{\alpha}%
}-i\theta_{i}^{\alpha}\sigma_{\alpha\dot{\alpha}}^{a} \pa_a
+{\bar{\theta}}_{i{\dot{\alpha}}}\pa_{\bar z}\right)  {\phi}\ ,\\[3mm]
D_z{\phi}={\partial}_{z}{\phi}\ ,  &
D_{\bar{z}}{\phi}={\partial}_{\bar{z}}{\phi}\ .
\end{array}
\label{coder}%
\end{equation}
and obey the same (anti)commutation rules as the generators, up to the sign
of the right-hand sides:
\begin{equation}
\left(  D_{A},D_{B}\right\}  =-T_{AB}^{C}D_{C}\ ,\label{al-coder}%
\end{equation}
the torsion coefficients $T_{AB}^C$  being given in (\ref{torsions}).

The components of the supermultiplet corresponding to the superfield ${\phi}$
are the coefficients of its expansion in powers of $\theta$ and ${\bar{\theta
}}$. A generic component can be written as
\begin{equation}
C_{n}=\left.  (D)^{n}{\phi}\right|  \ ,\label{gen-comp}%
\end{equation}
where $(D)^{n}$ is some product of $D_{\alpha}^{i}$ and ${\bar{D}}%
_{i{\dot{\alpha}}}$ derivatives, and where the symbol $|$ means that the
expression is evaluated at $\theta$ $=$ ${\bar{\theta}}=0$. It follows
from the explicit transformation rules
(\ref{transf-s-field}), that the action of the supersymmetry generators on the
components can be written as~\cite{soh}
\begin{equation}
Q_{a}^{i}C_{n}=\left.  D_{\alpha}^{i}(D)^{n}{\phi}\right|  \ ,
\quad{\bar{Q}}_{i{\dot{\alpha}}}C_{n}=
\left.D_{i{\dot{\alpha}}}(D)^{n}{\phi}\right|
\ ,\label{comp-transf}%
\end{equation}

\section{The generalized central charge constraint for the
free $N=2$ Fayet hypermultiplet}\label{cc-constr-free}

The Fayet hypermultiplet
\begin{equation}
\phi_{i}\equiv(\phi_{i},\chi_{\alpha},\bar{\psi}_{\dot{\alpha}},F_{i}%
)\label{Fayet}
\end{equation}
is formed by two SU(2) doublets of Lorentz scalars (${\phi}_{i},F_{i})$ 
and two
SU(2) singlets of Weyl spinors $\bar{\psi}_{\dot{\alpha}}$,
$\chi_{a}$. It will
represent the matter sector of an N=2 supersymmetric Yang-Mills theory but, 
in the present section, we shall only consider a brief
review of our previous work~\cite{I} on the free Fayet
hypermultiplet. The latter is defined by an SU(2)
 doublet
superfield\footnote{The same symbol ${\phi}$ represents the multiplet
(\ref{Fayet}), the corresponding superfield, as well as the first component of
the latter, i.e. its value at $\theta$ $=$ ${\bar{\theta}}$ $=0$.} ${\phi}%
_{i}({X})$ subjected to the supersymmetric constraints~\cite{fay,West}
\begin{equation}
D_{\alpha}^{i}\phi^{j}+D_{\alpha}^{j}\phi^{i}=0\ ,\qquad
 \bar{D}_{i\dot{\alpha}}
\phi^{j}+\bar{D}_{\dot{\alpha}}^{j}\phi_{i}=0\ .\label{Fayet-constr}
\end{equation}

\subsection{Central charge constraints and supersymmetry 
transformations}\label{cc-const-susy}

However, in order to define a finite supersymmetry representation, 
one has to impose a constraint which restricts the dependence of the 
superfield
$\phi_{i}(\bar{\phi}^{i})$ on the central charge coordinates $z$ 
and $\bar{z}$.
The constraint introduced in~\cite{I} reads
\begin{equation}
(\partial _{z} - \partial _{\bar{z}})\phi _{i}
=i\Lambda \phi _{i}\ ,\qquad
(\partial_{\bar{z}} - \partial _{z})\bar{\phi}^{i}
= i\Lambda\bar{\phi}^{i}  \ , 
\label{cc-constr}\end{equation}
It is a generalization of Sohnius'one~\cite{soh}, introducing a new
real parameter $\LA$ of dimension of a mass. 
\vspace{3mm}

\noindent{\bf Remark.} The constraint actually considered 
in~\cite{I} had $\LA$ complex\footnote{Note 
that the present parameter $\LA$ corresponds to $-i$ times the
parameter $\la$ of~\cite{I}.} and 
involved a factor $\exp(-v)$, with $v$ a
real parameter, in front of the $\bar z$ derivative in the first equation 
(and of the $z$ derivative in the second one). However it turned up
that the existence of an invariant action implied the reality
of $\LA$ and the vanishing of $v$.
This means that  we would not be able to
get a dynamics if $\LA$ were complex or
$v$ nonvanishing, although we still would
have a representation of the superalgebra (\ref{s-alg}). 
The reader may see~\cite{I} for more details.
\vspace{3mm}

Since the covariant derivatives $D$ and $\bar D$
\equ{coder} commute with $\partial_{z}$ and $\partial_{\bar{z}}$, 
the constraints above
hold for the superfield $\phi_{i}(\bar{\phi}^{i})$ and all its
derivatives, in particular the derivatives which define the component
fields \equ{gen-comp}.

Having defined the components of the hypermultiplet by
\begin{equation}%
\begin{array}[c]{l}%
{\phi}_{i}\equiv\left.  {\phi}_{i}\right|  \ ,\quad\chi_{a}\equiv\left.
\frac{1}{2\sqrt{2}}D_{\alpha}^{i}\phi_{i}\right|  \ ,\quad{\bar{\psi}}%
_{{\dot{\alpha}}}\equiv\left.  \frac{1}{2\sqrt{2}}{\bar{D}}_{{\dot{\alpha}}%
}^{i}{\phi}_{i}\right|  \ ,\quad F^{j}\equiv\left.  \frac{i}{8}D^{i\alpha
}D_{i\alpha}\phi^{j}\right|  =\left.  \partial_{z}\phi^{j}\ \right|  .
\end{array}
\label{fayet-comp}%
\end{equation}
we found the following $\LA$-dependent supersymmetry and
central charge transformation laws:
\begin{equation}
\begin{array}{ll}
Q_{\alpha }^{i}\phi _{j}=\sqrt{2}\delta _{j}^{i}\chi _{\alpha }\ ,\quad & 
\bar{Q}_{i\dot{\alpha}}\phi ^{j}=-\sqrt{2}\delta _{i}^{j}\bar{\psi}_{\dot{%
\alpha}}\ , \\[3mm] 
Q_{\a}^{i}\chi _{\b }=-\sqrt{2}i\varepsilon _{\a \b }F^{i}\ ,
& \bar{Q}_{i\dot{\alpha}}\chi _{\alpha }=-\sqrt{2}i\partial _{\alpha \dot{%
\alpha}}\phi _{i}\ , \\[3mm] 
Q_{\a }^{i}\bar{\psi}_{\dot{\beta}}=\sqrt{2}i\partial _{\a \dot{\beta}%
}\phi ^{i}\ , & \bar{Q}_{i\dot{\alpha}}\bar{\psi}_{\dot{\beta}}=
\sqrt{2}i\varepsilon _{\dot{\alpha}\dot{\beta}}(F_{i}-i\Lambda \phi _{i})\ ,
\\[3mm] 
Q_{\alpha }^{i}F_{j}=\sqrt{2}\delta _{j}^{i}\left( \partial _{\alpha 
\dot{\alpha}}\bar{\psi}^{\dot{\alpha}}+i\Lambda \chi _{\alpha }\right) \
,\quad & \bar{Q}_{i\dot{\alpha}}F^{j}=-\sqrt{2}\delta _{i}^{j}\partial
_{\alpha \dot{\alpha}}\chi ^{\alpha }\ ,
\end{array}
\label{Qtransfs}\end{equation}

\eq\ba{ll}
Z \phi ^{i} = F^{i}\ ,
  &\bar{ Z} \phi^{i} = (F^i - i\Lambda \phi^{i})\ , \es
Z \chi _{\alpha } = \partial _{\alpha \dot{\beta}}\bar{%
\psi}^{\dot{\beta}}+i\Lambda \chi _{\alpha }\ ,\qquad
  &\bar{ Z} \chi _{\alpha } =
       \partial_{\alpha \dot{\beta}}\bar{\psi}^{\dot{\beta}}  \ ,\es
Z \bar{\psi}_{\dot{\beta}} = 
    \partial_{\alpha \dot{\beta}}\chi^{\alpha }\ ,
  &\bar{ Z} \bar{\psi}_{\dot{\beta}}= (\partial
   _{\alpha \dot{\beta}}\chi ^{\alpha }-i\Lambda \bar{\psi}_{\dot{\beta}})\ , 
\es
Z F^{i} = \square \phi ^{i} + i\Lambda F^{i} \ ,
  &\bar{ Z} F^{i}=\square \phi ^{i}\ ,
\ea\eqn{Ztransforma}
and similarly for the conjugate components.
The algebra of these transformations closes as a representation of the
superalgebra \equ{s-alg}.
\vspace{3mm}

\subsection{The free Fayet Lagrangian}\label{free-lag}

The construction of an invariant Lagrangian in~\cite{I} is based
on a proposition due to Hasler~\cite{Hasler}:
\vspace{3mm}

\begin{enumerate}\item[]
\noindent{\bf Proposition}. 
Let be a superfield polynomial $L^{ij}$ 
-- called the ``kernel'' -- satisfying the conditions of zero symmetric
derivatives
\begin{equation}
D_{\alpha }^{(i}L^{jk)}=0\ , \qquad
\bar{D}_{\dot{\alpha}}^{(i}L^{jk)}=0\ .
\label{Hasler-cond}\end{equation}
Then the superfields
\begin{equation}
L\equiv -D_{k}^{\alpha }\Lambda _{\alpha }^{k}\ ,\qquad 
\bar{L}\equiv -{\bar D}_{k\dot{\alpha}}\bar{\Lambda}^{k\dot{\alpha}}\   ,
\label{Hasler-L}\end{equation}
where
\[
\Lambda _{\alpha }^{k}\equiv D_{i\alpha }L^{ik}\ ,\qquad 
\bar{\Lambda}^{k\dot{\alpha}}\equiv \bar{D}_{i}^{\dot{\alpha}}L^{ik} \ , 
\]
transform under supersymmetry -- with infinitesimal parameters 
$\xi$, $\bar\xi$ -- as
\begin{equation}\ba{l}
\delta L = i\pa_{z}
\lp \xi_i^\a\Lambda^i_\a+{\bar\xi}_{i\da}{\bar\Lambda}^{i\da}\rp
   -2i\partial_{a}
\lp{\bar\xi}_{i\da}{\bar\sigma}^{a\,\da\b}\Lambda^i_{\b}\rp\ ,\es
\delta \bar{L}=-i\pa_{\bar{z}}
\lp \xi_i^\a\Lambda^i_\a+{\bar\xi}_{i\da}{\bar\Lambda}^{i\da}\rp
   -2i\pa_{a}\lp \xi_i^\a \sigma^{a}_{\a\db}{\bar\Lambda}^{idb}\rp\ .
\ea\label{deltaL}\end{equation}
\vspace{3mm}
\end{enumerate}
Let us apply this proposition to the kernel used in~\cite{I}, 
\begin{equation}
L^{ij} = \pa_{\bar{z}}\bar{\phi}^{j}\phi^{i}
+ \bar{\phi}^{i} \pa_{z}\phi^{j} \ .
\label{kernel-k}\end{equation}
The central charge constraint \equ{cc-constr} implies that
$(\pa_x-\pa_{\bar z})L^{ij}$ is a total space-time derivative. Then the
free Lagrangian defined by
\eq
\LL_{\rm free} = \dfrac{-1}{24} (L + \bar L)\ ,
\eqn{free-lagr}
$L$ being defined by \equ{Hasler-L}, is 
indeed supersymmetry invariant up to a 
total space-time derivative.
Explicitly\footnote{See 
\equ{mult-conv} for the summation conventions.}:
\eq\ba{l}
\LL_{\rm free} = \bar{F}F- \pa_{a}\bar{\phi}\pa^{a}\phi 
 - i\chi\sla\pa\bar{\chi} -i\psi\sla\pa\bar{\psi} 
- i\frac{\LA}{2} \lp \bar F\f +i\bar\chi\bar\psi 
   + i\psi\chi - \bar \f F  \rp \ ,
\end{array}
\label{l-cin-l-lambda}
\end{equation}
with the notation $(\sla\pa\bar\psi)_\a$ $=$ 
$\pa_{\a\db} {\bar\psi}^\db$ $=$
$\s^a_{\a\db}\pa_a{\bar\psi}^\db$.

The terms in $\LA$ are mass terms, which have been induced
from the supersymmetry transformation rules written above. 
Writing down the equations of motion~\cite{I} indeed shoes that
$\LA/2$ represents the value of the mass of
the propagating fields $\f^i$,
$\chi$ and $\psi$, the field $F^i$ being auxiliary.

Of course, it is still possible to add a mass term ``by hand''. 
This can be done
with the help of Hasler's proposition, too, and leads to the independent 
invariant  mass Lagrangian 
\eq  
\LL_\m = \m \lp \bar{F}\phi + \bar{\phi}F + i\bar{\chi}\bar{\psi} 
 - i\psi \chi \rp \ ,
\eqn{mass-lagr}
where $\m$ is a real mass parameter.

However, imposing invariance under the parity 
transformations\footnote{The 4-component Dirac spinor
$\Psi=(\chi_\a,\,\bar\psi{}^\ad)$ then
transforms as $\Psi$ $\to$ $\g^0\Psi$.}
\eq\ba{l}
(x^0, {\bf x}) \rightarrow   (x^0, -{\bf x})\ ,\es
(\,\f^i\,,\         \chi^\a\,,\ \bar\psi{}_\db\,,\ F^i\,)
\leftrightarrow
(\,\bar\f{}_i\,,\ \bar\chi_\da\,,\ \psi{}^\b\,,\ -\bar F_i\,)\ ,
\ea\eqn{parity}
rules out the mass Lagrangian \equ{mass-lagr}.
Thus parity invariance insures that the mass is completely
determined by the parameter $\LA$.

\section{Coupling with an $N=2$ gauge supermultiplet}\label{gauge-fields}

\subsection{Gauge transformations and covariant derivatives in superspace}

The construction of N=2 supersymmetric Yang-Mills theory is based on a
SU(2) doublet of Fayet
hypermultiplets of matter fields, described by the superfields 
$\f^i$, $i=1,2$ and  now belonging to some representation 
$R$ of a compact Lie group $G$, the gauge group.
The conjugate superfield field $\f_i$ 
belongs to the conjugate representation $\bar R$. These superfields 
are subjected to a generalization of the constraints shown in the 
previous section, and  which will be introduced in the next subsection.
The generators of $G$ in the representation $R$
are antihermitean  matrices $T_r$  obey the Lie algebra commutation rules
\begin{equation}
\left[  T_{r},T_{s}\right]  =f_{rs}{}^{t}T_{t}\ .
\label{lie-alg}\end{equation}
 Local gauge invariance requires the 
introduction of a gauge connection superfield
\eq
\Phi_{A} = \Phi^r_{A}T_r\ ,\quad A= a,\, i\,\a,\, i\,\da,\,z,\,\bar z \ . 
\label{g-conn}\end{equation}
with the antihermicity conditions
\eq
(\F^a)^+ = -\F^a\ ,\quad (\F^i_\a)^+ = -\F_{i\da}  \ ,\quad 
(\F_z)^+ = +\F_{\bar z} \ .
\eqn{herm}
The infinitesimal gauge transformations read
\begin{equation}
\delta\phi^i=\Omega\phi^i\ ,\quad
   \delta\bar\phi{}_i =  \bar\f{}_i\Omega\ ,\quad
\delta\Phi_A = D_{A}\Omega + \lc\Omega, \Phi_{A}\rc\ ,
\label{g-transf}\end{equation}
where the  infinitesimal parameter $\Omega$ =  $\Omega^r T_r$ 
is an antihermitean superfield, subjected to some restrictions 
because of the constraints on the Fayet superfield $\f^i$, as it will be
shown in the next subsection. 
The covariant derivatives
\begin{equation}
\DD_A \phi^i = D_A \f^i-\Phi_A \phi^i\ ,\quad
\DD_A \bar{\phi}_i = 
   D_A\bar{\phi}_i + \bar{\phi}_i\Phi_A\ ,
\eqn{cov-dir}
with $D_A$ the ordinary superspace covariante derivative \equ{coder},
transform in the same way as $\f$ and $\bar\f$ in \equ{g-transf}.
Note that $\bar\f^i\f_i$ being gauge invariant, we have
\eq
\DD_A \lp \bar\f^i\f_i \rp = D_A \lp \bar\f^i\f_i \rp\ .
\eqn{cov-der-inv}
The super Yang-Mills curvature~\cite{Wess} 
\begin{equation}
\FF_{AB}=T_{AB}^{C}\Phi_{C}+D_{A}\Phi_{B}-(-)^{bc}D_{B}\Phi
_{A}-\left(  \Phi_{A},\Phi_{B}\right\}  \ ,
\label{curvature}\end{equation}
where the  torsion coefficients $T_{AB}^{C}$ are the same as in
the free case \equ{torsions},
transforms covariantly, in the adjoint representation:
\eq
\d\FF_{AB} = \lc \OM ,\FF_{AB} \rc\ .
\eqn{tr-curv}
Its covariant derivative is
\eq
\DD_A \FF_{BC} = D_A \FF_{BC} - \lp \F_A ,\FF_{BC} \rac\ .
\eqn{c-der-ad}
where we recall that the symbol $(\cdot,\,\cdot\}$ is an anticommutator 
if both entries are fermionic, and a commutator otherwise.
The commutation rules of the gauge covariant superspace 
derivatives \equ{cov-dir} or \equ{c-der-ad}
yields the supercurvature, e.g.:
\begin{equation}
\left(  \DD_{A},\DD_{B}\right\} \F_D  =
-T_{AB}^{C} \DD_{C}\, \F_D  - \lp\FF_{AB},\F_D \rac \ .
\label{algebra2}\end{equation}

\subsection{The gauge supermultiplet}

As usual~\cite{Wess,West,gsw}, the gauge superfields $\F_A$ must be
constrained in order to get a sensible gauge theory. In the present
case of central charge superspace, the natural constraints
consist in the vanishing
of all the supercurvature components with spinor 
indices~\cite{gai,hin,ghh}:
\begin{equation}
\FF_\a^i{}_\b^j =\FF_\da^i{}_\db^j
=\FF_\a^i{}_\db^j=0
\label{curv-cond}\end{equation}
Adding the constraint of central charge independence for the gauge
superfields:
\eq
\pa_z\F_A = \pa_{\bar z}\F_A = 0\ ,
\eqn{ccc-gfield}
one can use the Bianchi identities~\cite{West}
\begin{equation}
\dsum{{\rm cyclic}\,(ABC)}{}
\left(  \DD_{A}\FF_{BC}-T_{AB}%
^{E}\FF_{CE}\right)  =0\ ,
\label{Bianchi}\end{equation}
in order to show~\cite{hin} that all the nonvanishing 
curvature components may be expressed in
terms of the gaugino superfield $\F_\a^i$ and its conjugate:
\eq\ba{ll}
\FF^i_\a\,^a = i\s^a_{\a\da} \F^{i\da}\ ,\quad
&\FF_i^\da\,^a = -i{\bar\s}^{a\da\a} \F_\a^i\ ,\es
\FF_z\,_i^\da = -4 \F_i^\da\ ,\quad
&\FF_{\bar z}\,^i_\a = 4 \F^i_\a\ ,\es
\FF_{a\,b} = \dfrac{i}{16}\lp \s_{ab}^{\a\b}\DD_{\a\b}
  + {\bar\s}_{ab}^{\da\db}\DD_{\da\db} \rp \ ,\quad 
&\mbox{with}\quad 
\DD_{\alpha\beta} \equiv \left(  \DD_{\alpha}^{k}\DD_{k\beta}
+\DD_{\beta}^{k}\DD_{k\alpha}\right)\ ,  \es
\FF_{z\,\bar z} = -\lc \f_z,\f_{\bar z} \rp &\ .
\ea\eqn{curv(g-ino)}
Gaida~\cite{gai} has shown that all curvature components only depend on
the components $\F_z$ and $\F_{\bar z}$
of the gauge connection, through the identities
\eq
\F^i_\a = - \DD^i_\a \F_{\bar z}\ ,\quad
 \F_i^\da = -\DD_i^\da \F_z
\eqn{g-ino(X)}
In order to keep consistence with the condition \equ{ccc-gfield},
we must take the infinitesimal parameter $\Omega$ of
the gauge transformations \equ{g-transf} independent of $z$ and $\bar
z$. We then note that $\F_z$ and $\F_{\bar z}$
transform covariantly under the gauge transformations. Moreover,
due to \equ{Bianchi} and \equ{curv(g-ino)}, 
they obey to the covariantized chirality conditions 
\begin{equation}
\DD_{\dot{\alpha}}^{i}\F_{\bar z} = 0\ ,\quad 
\DD_{\alpha}^{i}\F_z = 0\ .
\label{chiral}\end{equation}
Supersymmetry transformations of the chiral superfield 
$\F_{\bar z}$ generate the
 components of the gauge supermultiplet, which consists of
a scalar $X$, an SU(2) doublet, spinor $X_{\alpha}^{i}$ (the gaugino), an
SU(2) triplet $X^{(ij)}$ and a Lorentz triplet $X_{\alpha\beta}$,
defined by
\eq\ba{l}
X = -\left. \F_{\bar z}\right|\ ,\es
X_\a^i = -\left. \DD_\a^i \F_{\bar z} \right|\ ,\es
X^{(ij)} = -\left. \DD^{ij} \F_{\bar z} \right|\ ,\quad \mbox{with}\quad
\DD^{ij} \equiv \DD^{i\a}\DD^j_\a + \DD^{j\a}\DD^i_\a \ ,\es
X_{(\alpha\beta)} = -\left. \DD_{\alpha\beta}\F_{\bar z} \right|\ ,\quad
\mbox{with}\quad 
\DD_{\alpha\beta} \equiv \left(  \DD_{\alpha}^{k}\DD_{k\beta}
+\DD_{\beta}^{k}\DD_{k\alpha}\right)\ .
\ea\eqn{g-mult-comp}
The SU(2) triplet obeys the condition of reality:
\[
(X^{ij})^+= - X_{ij})^+ \ .
\]
The Lorentz triplet and its conjugate are linked to the Yang-Mills
curvature 
\[
\FF_{ab} = \pa_a A_b - \pa_b A_a - \lc A_a,A_b \rc\ ,\quad
\mbox{where }\ A_a = \left.\F_a\right|
\]
by:
\[
X_{\alpha\beta}=8i\sigma_{\alpha\beta}^{ab}\FF_{ab}\ ,\quad
\bar{X}_{\dot{\alpha}\dot{\beta}}
=8i{\bar\sigma}_{\dot{\alpha}\dot{\beta}}^{ab}\FF_{ab}\ ,
\]
or, conversely:
\eq
\FF^{ab}\ =-\frac{i}{16}(\sigma_{\alpha\beta}^{ab}X^{\alpha\beta} 
+\sigma_{\dot{\alpha}\dot{\beta}}^{ab}\bar{X}^{\dot{\alpha}\dot
{\beta}})\ ,
\eqn{LorentzCurvature}
The transformation laws of the gauge supermultiplet are obtained using
the definition \equ{g-mult-comp} of the components and Equation
\equ{comp-transf} with the ordinary superspace 
spinor derivatives $D_\a^i$, $\bar D{}_\da^i$  
replaced by the covariant derivatives $\DD_\a^i$, $\DD{}_\da^i$.
The result is:
\begin{equation}\begin{array}{l}
Q_{\alpha}^{i}X = X_{\alpha}^{i}\es
Q_{\alpha}^{i}X_{\beta}^{j}=-i\varepsilon^{ij}\varepsilon
_{\alpha\beta}\left[  \bar{X},X\right]  +\frac{1}{4}\varepsilon_{\alpha\beta
}X^{ij}-\frac{1}{4}\varepsilon^{ij}X_{\alpha\beta}\es
Q_{\alpha}^{i}X^{jk}=4i\DD_{\alpha\dot{\beta}}%
(\varepsilon^{ij}\bar{X}^{\dot{\beta}k}+\varepsilon^{ik}\bar{X}^{\dot{\beta}%
j})-4i\left[  \varepsilon^{ij}X_{\alpha}^{k}+\varepsilon^{ik}X_{\alpha}%
^{j},\bar{X}\right]  \es
Q_{\alpha}^{i}X_{\beta\gamma}=4i(\varepsilon_{\alpha\beta}
\DD_{\gamma\dot{\alpha}}+\varepsilon_{\alpha\gamma}\DD%
_{\beta\dot{\alpha}})\bar{X}^{i\dot{\alpha}}\es
\bar{Q}_{k\dot{\alpha}}X=0\es
\bar{Q}_{k\dot{\alpha}}X_{\beta}^{j}
=-2i\delta_{k}^{j}\DD_{\b\dot{\alpha}}X \es
\bar{Q}_{k\dot{\alpha}}X^{ij}=4i\DD_{\beta\dot{\alpha}}(\delta
_{k}^{i}X^{j\beta}+\delta_{k}^{j}X^{i\beta})-4i\left[  \delta_{k}^{i}\bar
{X}_{\dot{\alpha}}^{j}+\delta_{k}^{j}\bar{X}_{\dot{\alpha}}^{i},X\right]
\es
\bar{Q}_{k\dot{\alpha}}X_{\alpha\beta}=-4i(\DD_{\alpha\dot{\alpha
}}X_{k\beta}+\DD_{\beta\dot{\alpha}}X_{k\alpha})
\end{array}
\label{transf-chiral}\end{equation}
and similar transformations for the conjugated multiplet $\bar
{X}=(\bar{X},\bar{X}_{i\dot{\alpha}},{X}_{ij}\bar{X}_{\dot{\alpha}%
\dot{\beta}})$
In the equations above, 
$\DD_{\beta\dot{\alpha}}\equiv \sigma_{\beta\dot{\alpha}}^{a}\DD_{a}$
= $\s_{\b\da}^{a}(\pa_a-[A_a,\ \,])$.

\subsection{Gauge Lagrangian}

The gauge field supermultiplet being chiral, the corresponding
gauge invariant Lagrangian, supersymmetric up to a total derivative,
may be defined as
\begin{equation}
\LL_{\rm gauge}=\frac{1}{3\cdot2^9} \tr \lp  \DD^{ij}\DD_{ij}%
(\F_{\bar z})^{2} + \mbox{c.c.} \rp \ ,\label{Lgauge}%
\end{equation}
with $\DD^{ij}$ defined in (\ref{g-mult-comp}). Using
the transformations laws (\ref{transf-chiral}) and equation
(\ref{LorentzCurvature}) we explicitly get
\begin{equation}\begin{array}{l}
\mathcal{L}_{\mathrm{gauge}}
=
\frac{1}{4}\mathrm{{Tr}\,}\mathcal{D}^{a}\mathcal{D}_{a}%
\bar{X}X
+\frac{1}{4}\bar{X}\mathcal{D}_{a}\mathcal{D}^{a}X
-\frac{i}{8}(\mathcal{D}%
_{\alpha \dot{\alpha}}\bar{X}^{i\dot{\alpha}})X_{i}^{\alpha }
+\frac{i}{8}%
\bar{X}^{\dot{\alpha}i}(\mathcal{D}_{\alpha \dot{\alpha}}X^{i\alpha }) \es
-\frac{1}{4}\FF_{ab}\FF^{ab}
+\frac{1}{256}X^{ij}X_{ij}
+\frac{i}{8}\bar{X}\{X^{\alpha i},X_{\alpha i}\}
+\frac{i}{8}\{\bar{X}_{\dot{%
\alpha}}^{i},\bar{X}_{i}^{\dot{\alpha}}\}X
+\frac{1}{8}[X,\bar{X}]\bar{X}X\ .
\end{array}
\label{Lgauge2}
\end{equation}

\subsection{Generalized Fayet-Sohnius constraints. 
Supersymmetry transformations of the hypermultiplet}

The natural extension of the condition (\ref{Fayet-constr}) defining 
$\phi^{i}$ as a Fayet hypermultiplet, in the presence of a gauge
connection, is given by:
\begin{equation}
\DD_{\alpha}^{i}\phi^{j}+\DD_{\alpha}^{j}\phi^{i} 
=0 \ ,\quad
 \DD_{\dot{\alpha}}^{i}\phi^{j}
+\DD_{\dot{\alpha}}^{j}\phi^{i}=0\ ,
\label{cov-fay-constr}\end{equation}
and the conjugate equations. The use of the covariant derivative
guarantees the compatibility of this condition with gauge covariance.

For what concerns the generalized central charge constraint on
the Fayet superfield, necessary in order to 
get a finite supermultiplet, we shall use the same condition 
\equ{cc-constr} as in the free case:
\begin{equation}
(\partial _{z} - \partial _{\bar{z}})\phi _{i}
=i\Lambda \phi _{i}\ ,\qquad
(\partial_{\bar{z}} - \partial _{z})\bar{\phi}^{i}
= i\Lambda\bar{\phi}^{i}  \ , 
\label{Zconstraint}\end{equation}
The compatibility of this constraint with gauge covariance relies on the
fact that the $z$ and $\bar z$ components of the connection, namely
$\bar X$ and $x$ (see \equ{g-mult-comp}),
 are covariant, which implies the covariance of 
the partial derivatives $\pa/\pa z$ and $\pa/\pa\bar z$. Let us mention
that a slightly more general condition is possible here, too,
as we have noted in
Section \ref{cc-constr-free} for the free case, but which as well in the
present interactive case would prejudicate the existence of an
invariant action.

The definition of the hypermultiplet components in the presence of the 
gauge connection is a covariant
extension of the definition (\ref{fayet-comp}) proposed in the free case:
\begin{equation}\begin{array}{l}
{\phi}_{i} = \left.  {\phi}_{i}\right|  \ ,\quad\chi_{a} = \left.
\frac{1}{2\sqrt{2}}\DD_{\alpha}^{i}\phi_{i}\right|  \ ,\quad
{\bar{\psi}}_{{\dot{\alpha}}} = \left.  \frac{1}{2\sqrt{2}}
{\bar\DD}_{{\dot{\alpha}}}^{i}{\phi}_{i}\right|  \ ,\es
F^{j}  =\left.  \partial_{z}\phi^{j}\ \right| 
= \left.\lp\frac{i}{8}\DD^{i\alpha}\DD_{i\alpha}\phi^{j}
+\Phi_{z}\phi^{j} \rp\right|
\end{array}
\label{cov-comp}%
\end{equation}
and similarly for the conjugated hypermultiplet $\bar{\phi}^{i}$.
Let us note that the component $F^i$ is defined as the
simple partial derivative  $\pa/\pa z$ instead of
the covariant one. The latter and its conjugate then read, at
$\theta=0$, 
\begin{equation}\ba{l}
\left.\DD_{z}\phi^{i}\right|
=\left.(\partial_{z}-\Phi_{z})\phi^{i}\right| =
 F^{i}+\bar{X}\phi^{i}\ ,\es
\left.\DD_{\bar{z}}\phi^{i}\right|
=\left.(\partial_{\bar{z}}-\F_{\bar z})\phi^{i}\right|
= F^{i} +(X-i\Lambda)\phi^{i} 
\ea\label{DZ}\end{equation}
The supersymmetry transformations of the components are then defined by
the covariantized expressions \equ{comp-transf}.
Using the definitions above, the curvature conditions (\ref{curv-cond}), 
the Fayet constraints (\ref{cov-fay-constr}) and the central charge
constraints \equ{Zconstraint} we find, with the help of the Bianchi
identities \equ{Bianchi}, the following supersymmetry
transformation laws of the hypermultiplet:
\begin{equation}\ba{l}
\ba{ll}
Q_\a^i \phi_{j} = \sqrt{2}\delta_{j}^{i}\chi_{\alpha}\ ,\quad&
{\bar Q}_{i\da}\phi^{j} = -\sqrt{2}\delta_{i}^{j}\bar{\psi}_{\dot{\alpha}}\ ,\es
Q_\b^i \chi_{\alpha}=-\sqrt{2}i\varepsilon_{\beta\alpha
}(F^{i}+\bar{X}\phi^{i}) \ ,\quad&
{\bar Q}_{i\da}\chi_{\beta}=-\sqrt{2}i\DD_{\beta
\dot{\alpha}}\phi_{i}\ ,\es
Q_\b^i \bar{\psi}_{\dot{\beta}}=\sqrt{2}i\DD%
_{\beta\dot{\beta}}\phi^{i}\ ,\quad&
{\bar Q}_{i\da}\bar{\psi}_{\dot{\beta}}
=\sqrt{2}i\varepsilon_{\dot{\alpha}\dot{\beta}}
 \left( F_{i}-(i\Lambda-X)\phi_{i}\right)  \ ,
\ea\\[12mm]
\ba{l}
Q_\a^i F_{j}=\sqrt{2}\delta_{j}^{i}
\left(\DD_{\alpha\dot{\alpha}}\bar{\psi}^{\dot{\alpha}}
+(i\Lambda-X)\chi_{\alpha}-X_{\alpha}^{k}\phi_{k}\right)  \ ,\es
{\bar Q}_{i\da}F^{j}=-\sqrt{2}\delta_{i}^{j}\left(
\DD_{\a\da}\chi^{\alpha}-\bar{X}\bar{\psi}_{\dot{\alpha}%
}+\bar{X}_{\dot{\alpha}}^{k}\phi_{k}\right) \ .
\ea
\ea\label{Fayet-cov-transf}\end{equation}
In the same way one finds the central charge transformations
\begin{equation}\ba{l}
Z\phi^{i}=F^{i}+\bar{X}\phi^{i}\ ,\quad
 \bar Z\phi^{i}= F^{i} - (i\Lambda- X)\phi^{i}\, \es
Z\chi_{\alpha}= \DD_{\alpha\dot{\alpha}}\bar
{\psi}^{\dot{\alpha}}+(i\Lambda- X+\bar{X})\chi_{\alpha}- X_{\alpha
}^{k}\phi_{k} \ ,\quad
 \bar Z\chi_{\alpha}=\DD_{\alpha
\dot{\alpha}}\bar{\psi}^{\dot{\alpha}}-\frac{1}{\sqrt{2}}X_{\alpha}^{k}%
\phi_{k}\ ,\es
Z\bar{\psi}_{\dot{\alpha}}=\DD_{\alpha\dot{\alpha}}\chi^{\alpha
}+\bar{X}_{\dot{\alpha}}^{k}\phi_{k}\ ,\quad
 \bar Z\bar{\psi}_{\dot{\alpha}}= \DD%
_{\alpha\dot{\alpha}}\chi^{\alpha}+\bar{X}_{\dot{\alpha}}^{k}\phi
_{k}-(i\Lambda- X+\bar{X})\bar{\psi}_{\dot{\alpha}}\ , \es
ZF^{i}= \square\phi^{i} + (i\Lambda+\bar{X})F^{i} \ ,\quad
 \bar ZF^{i}=\square\phi^{i}+XF^{i} 
\ea
\label{GaugeCentralChargeTransf}\end{equation}
and similar transformations for the conjugated hypermultiplet $\bar{\phi}^{i}
$. In these equations, $\DD_{\a\dot{\alpha}} = \s_{\a\da}^a\DD_{a}$
= $\s_{\a\da}^{a}(\pa_a-A_a)$.

\subsection{The hypermultiplet Lagrangian minimally coupled
to the gauge connection}

The construction of the hypermultiplet Lagrangian minimally coupled
to the gauge connection follows the same lines as in the free case,
being
based on Hasler's proposition stated in Section~\ref{free-lag}. We first
observe that the kernel $L^{ij}$ (\ref{kernel-k}) is
gauge invariant as it reads. This way, we have
\begin{equation}
D_{A}L^{ij}=\DD_{A}L^{ij}\ ,\ \mbox{for all A}\ ,
\label{covderivkernel}\end{equation}
and so, the natural extension of Hasler's procedure\footnote{This
procedure is detailed in~\cite{I}.}
in order to get a gauge
invariant Lagrangian, supersymmetry invariant up to a total derivative,
is to
substitute in every step
the ordinary superspace derivative $D_A$ used in the free case, 
by the gauge covariant derivatives $\DD_A$,  using the property
\eq
D_A \tr (\bar\vf\vf') = \tr \lp \DD_A\bar\vf\vf' 
+ \bar\vf\DD_A\vf' \rp\ ,
\eqn{leibniz}
$\vf$ and $\vf'$ being the Fayet superfield or some
of its covariant derivatives.
Since the latter define the
components of the hypermultiplet and their transformation rules
through \equ{cov-comp} and the covariantized form 
of (\ref{comp-transf}), it is
easy to compute the Lagrangian using the supersymmetry transformation
laws \equ{Fayet-cov-transf}. 
Thus, starting with the kernel \equ{kernel-k}, we get
\begin{equation}\begin{array}{l}
\LL_{\rm hypermultiplet}=\bar{F}F
-i\psi^{\alpha}\DD_{\alpha\dot{\beta}}\bar{\psi}^{\dot{\beta}}
+i\bar{\chi}^{\dot{\beta}}\DD_{\alpha\dot{\beta}}\chi^{\alpha}
-\DD^{a}\bar{\phi}\DD_{a}\phi  \es
+i\psi X\chi-i\bar{\chi}\bar{X}\bar{\psi}
+\frac{i}{\sqrt{2}}\psi^{\alpha}X_{\alpha}^{k}\phi_{k}
-\frac{i}{\sqrt{2}}
  \bar{\phi}^{k}\bar{X}_{\dot{\alpha}k}\bar{\psi}^{\dot{\alpha}}
-\frac{i}{\sqrt{2}}\bar{\phi}^{k}X_{k\alpha}\chi^{\alpha}
-\frac{i}{\sqrt{2}}\bar{\chi}_{\dot{\alpha}}\bar{X}^{\dot{\alpha}k}\phi_{k} \es
-\frac{1}{2}\bar{\phi}(\bar{X}X+X\bar{X})\phi
+\frac{i}{8}\bar{\phi}^{i}X_{ij}\phi^{j} \es
-\frac{i}{2} \Lambda \lp \bar{F}\phi-\bar{\phi}F+
i\psi\chi+i\bar{\chi}\bar{\psi}
-\bar{\phi}\bar{X}\phi+\bar{\phi}X\phi \rp \ ,
\end{array}
\label{cov-fayet-lag}\end{equation}
As in the free case, an independent invariant mass Lagrangian may be 
added, in the form of the gauge invariant extension of the supersymmetric
mass Lagrangian
\equ{mass-lagr}:
\eq  
\LL_\m = \m \lp \bar{F}\phi + \bar{\phi}F + i\bar{\chi}\bar{\psi} 
 - i\psi \chi +\bar\f X\f + \bar\f\bar X\f    \rp \ ,
\eqn{g-inv-mass}
with $\m$ a real mass parameter. However this mass Lagrangian again 
is ruled out by the requirement of invariance under the parity
transformations
\eq\ba{l}
(x^0, {\bf x}) \rightarrow   (x^0, -{\bf x})\ ,\es
(\,\f^i\,,\         \chi^\a\,,\ \bar\psi{}_\db\,,\ F^i\,)
\leftrightarrow
(\,\bar\f{}_i\,,\ \bar\chi_\da\,,\ \psi{}^\b\,,\ -\bar F_i\,)\ ,\es
(\, X\,,\ \bar X\,,\ X_\a^k\,,\ {\bar X}^\da_k\,,\ X^{ij}\,)
\rightarrow
(\, -{\bar X}^{\rm T}\,,\ -X^{\rm T}\,,\ -({\bar X}^{\rm T})^\da_k\,,\ 
-(X^{\rm T})_\a^k\,,\ (X^{\rm T})_{ij}\,)
\ea\eqn{parity'}
where the superscript T means transposi\cao.

In order to complete the description of the model, we write down the
field equations for the matter fields:
\begin{equation}\ba{l}
F_{i} - iM\phi_{i}=0\ ,
\es
\DD_{a}\DD^{a}\phi_{i} + iM F_{i} 
+ \lp iM(\bar{X}-X) - \frac{1}{2}(\bar{X}X+X\bar{X}) \rp\phi_{i}
\es \qquad - \frac{i}{\sqrt2}
 (\bar X{}_{i\da}\bar\psi{}^\da+X_{i\a}\chi^\a)\phi_{i}
+\frac{i}{8}X_{ik}\phi^{k}=0\ ,
\es
-i\DD_{\alpha\dot{\beta}}\bar{\psi}^{\dot{\beta}}
+ (M+iX)\chi_{\alpha} + \frac{i}{\sqrt{2}}X_{\alpha}^{k}\phi_{k}%
=0\ ,
\es
i\DD^{\a\db}\chi_{\alpha} - (M+i\bar{X})\bar{\psi}^{\dot{\beta}}
-\frac{i}{\sqrt{2}}\bar{X}^{\dot{\beta}k}\phi_{k}%
=0
\ea\label{eq-motion}\end{equation}
where the mass $M$ is given by
\[
M=\dfrac{\LA}{2}\ .
\]

\section{Conclusions}

We have shown that the generalized central charge constraint
proposed in~\cite{I} for the free Fayet hipermultiplet
 is working in the case of minimal coupling with a super-Yang-Mills 
connection. In this case, too, the constraint modifies the 
supersymmetry transformation rules with terms depending of
a parameter having the dimension of a mass. This parameter
indeed shows up in the resulting action as the mass of the 
hypermultiplet. Moreover the mass is totally induced by this 
mechanism if parity invariance is imposed, which lets us conclude 
to the possibility of a nonrenormalization theorem for the mass.

\subsubsection*{Acknowledgments}

We are much grateful to Richard Grimm for his help and very interesting
discussions, in particular during his visit in Brazil, made possible by a
financial support of the CAPES/COPLAG, and during a month's stay of one of the
authors (O.P.) as Professor Invit\'{e} at the Centre de Physique Th\'{e}orique
(CPT) of the Universit\'{e} de Aix-Marseille, France. O.P. would like to
warmly thank the members of the CPT for their very kind invitation and for
their hospitality.

\section*{Appendix. Notations and conventions}

\renewcommand{\theequation}{\Alph{section}.\arabic{equation}} %
\setcounter{section}{1} 
\setcounter{equation}{0}

Space-time is Minkovskian, 4-vector components are 
labelled by latin letters $a,b,\cdots$ 
$=$ $0,1,2,3$, the metric is choosen as  
\eq
\eta_{ab} = \mbox{diag} (-1, 1, 1, 1) \ .
\eqn{metric}
Weyl spinors are complex 2-component spinors $\p_{\a}$, $\a=1,2$, in the
$({1\over2},0)$ representation of the Lorentz group, or 
$\p_{\da}$, $\da=1,2$, in the $(0,{1\over2})$ representation. The $N=2$
internal symmetry group is ``isospin'' SU(2), isospinors being denoted
by $X^i$, $i=1,2$.

Isospin indices $i$ are raised and
lowered by the antisymmetric tensors $\e^{ij}$ and $\e_{ij}$:
\eq\ba{l}
X^{i}=\e^{ij}X_{j}\ ,\quad X_{i}=\e_{ij}X^{j}\ , \es
\mbox{with: \ }
\e^{ij}=-\e^{ji}\ ,\quad \e^{12}=1\ ,\quad \e_{ij}\e^{jk}=\d_i^k\ ,\quad
\e^{ij}\e_{kl}=\d_{l}^{i}\d_{k}^{j}-\d_{k}^{i}\d_{l}^{j}\ .
\ea\eqn{isospin-ind}
The same holds for the Lorentz spin indices, with the tensors $\e^{\a\b}$ and  
$\e^{\da\db}$ obeying to the same rules \equ{isospin-ind}.

Multiplication of spinors and isospinors is done, if not otherwise stated,
according to the convention
\eq
\p\chi = \p^\a\chi_\a\ ,\quad 
\bar\p\bar\chi = {\bar\p}_\da{\bar\chi}^\da\ ,\quad
UV = U^i V_i\ . 
\eqn{mult-conv} 

Our conventions for the complex conjugation, denoted by $^*$, are as follows:
\eq
(X^i_\a)^* = {\bar X}_{i\da}\ ,\quad 
({\bar X}_{i\da})^* = X^i_\a\ .
\eqn{compl-conj}

The matrices $\s^a$ and ${\bar\s}^a$ are defined by  
\eq\ba{l}
{\bar\s}^{a\,\da\a} = \e^{\a\b} \e^{\da\db} \s^a_{\b\db}\ ,\es
\s^0 = \identity \ ,\quad 
 \s^i\,(i=1,2,3)\, = \mbox{Pauli matrices} \ , \es
\bar\s{}^0 = \identity \ ,\quad 
 \bar\s{}^i\,(i=1,2,3)\, = -\mbox{Pauli matrices} \ , 
\ea\eqn{Pauli-matrices}
and obey the properties  
\eq
\s^a{\bar\s}^b + \s^b{\bar\s}^a = - 2\eta^{ab}\ ,\quad
\s^a_{\a\da} {\bar\s}_a^{\db\b} = -2\d^\b_\a \d^\db_\da\ .
\eqn{prop-P-matr}


\end{document}